\documentclass[twocolumn,showpacs,preprintnumbers,amsmath,amssymb,superscriptaddress,prb]{revtex4}

\usepackage{graphicx}
\usepackage{dcolumn}
\usepackage{bm}
\usepackage{subfigure}
\usepackage{color}
\usepackage{amsfonts}%
\usepackage{amssymb}
\usepackage{amsmath}%
\usepackage{wasysym}

\begin{document}


\title{Terahertz Kerr and Reflectivity Measurements on the Topological Insulator Bi$_2$Se$_3$}

\author{G. S. Jenkins}
\author{A. B. Sushkov}%
\author{D. C. Schmadel}
\author{N. P. Butch}
\author{P. Syers}
\author{J. Paglione}
\author{H. D. Drew}
\affiliation{%
Center for Nanophysics and Advanced Materials\\Department of
Physics, University of Maryland, College Park, Maryland 20742, USA
}%

\date{\today}

\begin{abstract}

We report the first terahertz Kerr measurements on bulk crystals
of the topological insulator Bi$_2$Se$_3$. At T=10\,K and fields
up to 8\,T, the real and imaginary Kerr angle and reflectance
measurements utilizing both linearly and circularly polarized
incident radiation were measured at a frequency of 5.24\,meV. A
single fluid free carrier bulk response can not describe the
line-shape. Surface states with a small mass and surprisingly
large associated spectral weight quantitatively fit all data.


\end{abstract}

\maketitle

\section{Introduction}

The
discovery\cite{KaneMele_PRL_2007,moore_Balents_PRB2007,roy_PRB2009,FuKane_PRB2007,BernevigZhang_Science2006,BernevigZhang_PRL2006,QiZhang_PRB2008,Zhang_BiSeBiTeSbTe_2009}
and realization of 3D strong topological insulators
(TIs)\cite{Hsieh-Hasan-ARPES-BiSb-Nature2008,Hsieh-Hasan_ARPESspintexture-Science2009,Xia-Hassan_ARPESBiSe_2009,Hsieh-Hassan-ARPESBiSeSbTe_2009}
have generated much recent interest. A strong spin-orbit coupling
can produce a topologically protected metallic surface state which
has a Dirac-like dispersion with a non-degenerate chiral spin
state that can not be destroyed or gapped by (non-magnetic)
impurity scattering. Since the spin direction of the quasiparticle
is locked to the momentum
direction,\cite{Hsieh-Hasan_ARPESspintexture-Science2009}
suppression of Anderson localization results from the destructive
interference associated with the quaisparticle's Berry phase upon
backscattering.\cite{Yazdani_STM-Backscatteringsuppressed_2009}
With the action of local ordered magnetic moments, the surface
state becomes gapped.\cite{QiZhang_PRB2008} In zero magnetic field
with the Fermi energy within the gap, a quantized Kerr angle
associated with the half quantized Hall conductance is a predicted
signature of the topological origin of the surface
state.\cite{QiZhang_PRB2008}

These and other predicted properties have stimulated a recent
flurry of activity to probe the exotic surface state. Although the
surface states have been observed by angular resolved
photo-emission spectroscopy
(ARPES),\cite{Hsieh-Hasan-ARPES-BiSb-Nature2008,Hsieh-Hasan_ARPESspintexture-Science2009,Xia-Hassan_ARPESBiSe_2009,Hsieh-Hassan-ARPESBiSeSbTe_2009}
unambiguous separation of bulk from surface state conduction has
plagued transport characterization of the surface state due to the
rather large intrinsic charge density in the bulk conduction band
in as-grown Bi$_2$Te$_3$ and Bi$_2$Se$_3$. Various schemes to
circumvent this difficulty in bulk crystals have either
implemented counter doping during
growth,\cite{Ong_CaDopingAndTeDoping_2010} minimized Se
vacancies,\cite{butch_PRB2010} or fine tuned annealing
processes.\cite{Ong_CaDopingAndTeDoping_2010} One intermediate
goal has been to sufficiently deplete the bulk to allow gating or
chemical doping of the surface in suitably thin crystals to move
the chemical potential of the bulk into the
gap.\cite{Ong_ThinBiSeGating_2010}

There are several advantages of terahertz Kerr measurements over
other transport measurements which allow separation of the surface
from the bulk response. The surface states are expected to have a
a different mass than the bulk, depending upon the location of the
chemical potential, giving rise to two different resonant
cyclotron frequencies. This effect can be further exaggerated by
application of an electric field either from gating or chemically
doping the surface causing `band bending' effects, pushing the
Dirac point closer to the chemical potential. Secondly, the
line-shape as a function of magnetic field associated with a
surface state is qualitatively different than the bulk response.
Thirdly, no electrical contacts are required. Lastly, Kerr
measurements on thick crystals allow probing of a single surface
state.

In this article, Kerr measurements are reported on two bulk
crystals of Bi$_2$Se$_3$ at an optical frequency of 5.24\,meV in
fields up to 8\,T. Line-fits of the data show evidence of surface
states. For a measured surface state cyclotron mass of $m_{cs} =
0.033\, m_e$, the associated spectral weight is about a factor of
20 times larger than expected based upon the corresponding number
density estimated from ARPES data. Measurements of other samples
from the same as-grown crystal show a higher carrier
concentration. Alternatively, a simple model of the optical
response which assumes the presence of two crystallites with
different carrier concentrations can qualitatively describe all of
our data. Future work to optically characterize carrier density
homogeneity as well as an analysis of surface state signals
expected by gating are discussed.

\begin{figure}
  \centering
  \includegraphics[scale=1]{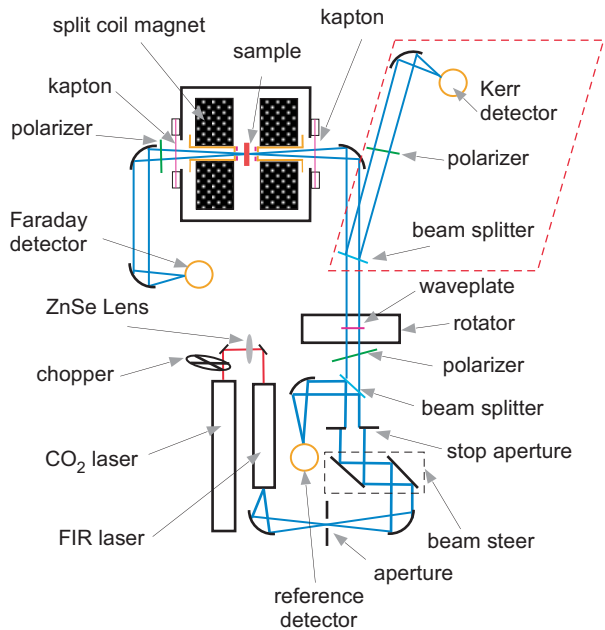}
  \caption[Diagram of Faraday experiment]{(Color online) Schematic of the Faraday and Kerr optical system
   capable of operating on various laser lines in the terahertz spectral region. The figure, adapted from
   Ref.\,\onlinecite{RSI-Jenkins} which details the Faraday experiment, depicts additional
   optics within the red-dashed trapezoid allowing Kerr measurements by capturing reflected light from the sample.
    }\label{figOptics}
\end{figure}

\begin{figure*}[!t]
\includegraphics[scale=.25,clip=true]{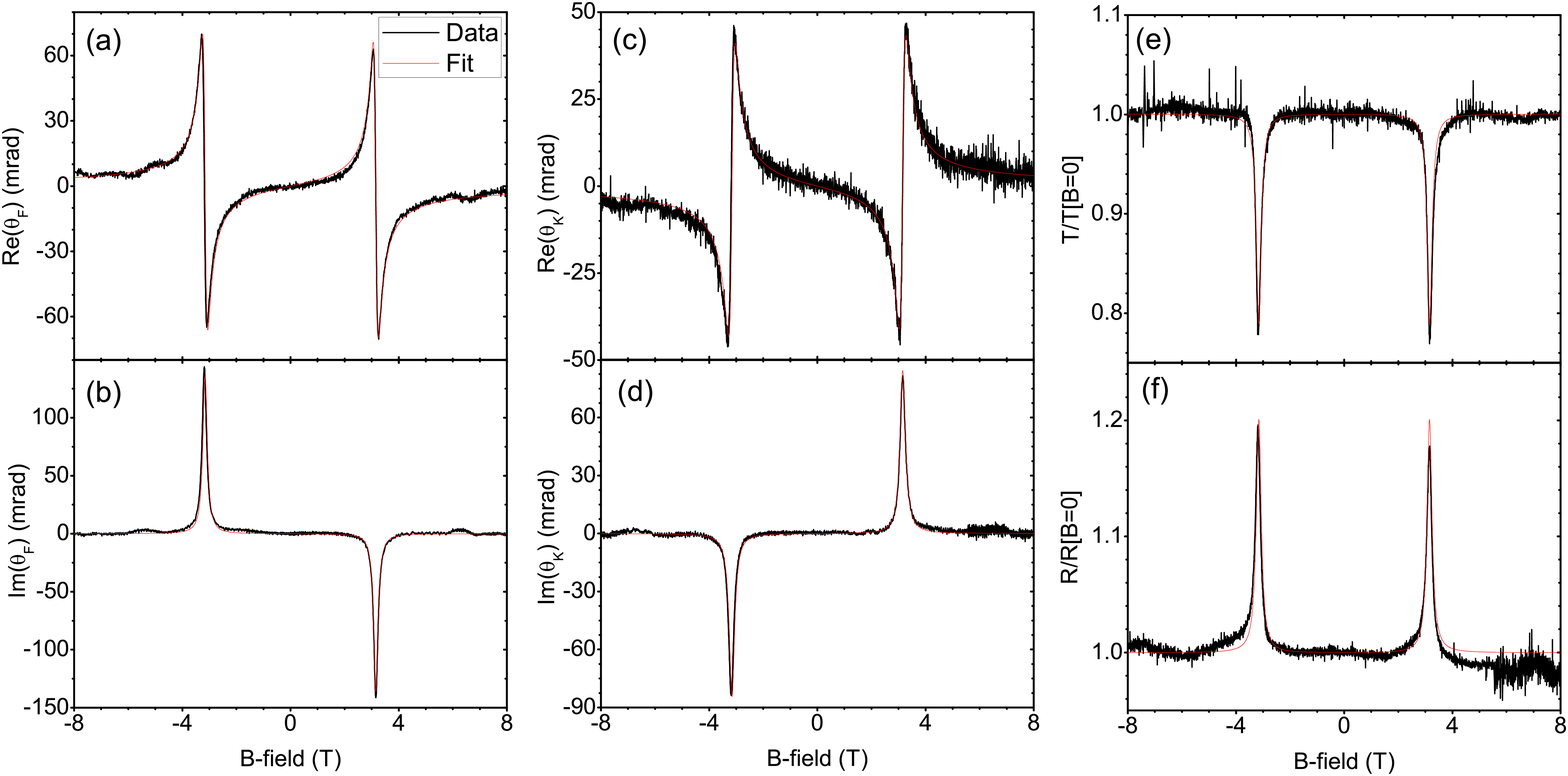}
\caption{\label{Cal2deg}(color online) Calibration of the Faraday
and Kerr angle measurement system. The Faraday angle, Kerr angle,
and relative transmission and reflection from a GaAs 2-DEG
heterostructure were measured at an optical frequency of 5.24 meV
and temperature of 80\,K as a function of magnetic field. The
Faraday and Kerr angles are simultaneously fit with effectively
two parameters, the number density and mobility, found to be
n=$2.0\times10^{11}\,\text{cm}^{-2}$ and $\mu=-1.19 \times 10^5 \,
\text{cm}^2 / \text{V\,sec}$. The fits (red line) are overlayed on
top of the data (black dots) for comparison. The same two
parameters are used to calculate reflection and transmission
(normalized to the B=0 value) and overlayed on the data.}
\end{figure*}

\section{Experimental Description}
\subsection{Kerr System Description and Calibration}

For calibration purposes, the Faraday rotation and ellipticity
induced by a sample in magnetic field was measured utilizing a
polarization modulation technique with a heterodyne detection
scheme, a technique detailed elsewhere.\cite{RSI-Jenkins} The
complex Faraday angle and relative transmission were concurrently
measured.

The instrument was modified allowing simultaneous measurement of
the complex Kerr angle and relative reflection by the insertion of
a beam splitter, polarizer, and collection optics which include an
off-axis parabolic aluminum mirror and a 2\,K silicon composite
bolometer as depicted in Fig.\,\ref{figOptics}. The Kerr and
Faraday angle are measured separately.

Assuming a rotationally invariant sample, the complex Kerr and
Faraday angle are defined as follows:
\begin{eqnarray}
\tan(\theta_K) &=  i (r_+ - r_-)/ (r_+ + r_-)\notag\\
\tan(\theta_F) &=  i (t_+ - t_-)/ (t_+ + t_-)\notag
\label{eq:FAKAdefs}
\end{eqnarray}
where r (t) is the reflection (transmission) complex Fresnel
amplitude of the sample expressed in a circular polarization
basis. The + and - represent right and left circular polarization
states, respectively.

Absolute measurement of the real part of the Faraday and Kerr
angles is verified by inserting a linear polarizer into the system
immediately downstream from the rotator. The measured rotation
angle equals the polarizer angle for small angles which is a very
good approximation for the signals measured in Bi$_2$Se$_3$. Large
angle calibration may be accomplished in a similar manner, but
corrections from higher order terms involving the Fresnel
amplitudes must be explicitly taken into account in the modulated
signal.\cite{RSI-Jenkins}

Calibration of the imaginary part of the Faraday and Kerr angles
is accomplished by measuring the known response of a GaAs 2-DEG
heterostructure. A NiCr antireflection (AR) coating was deposited
on the backside to minimize multiple reflections within the GaAs
substrate. There is a well defined relationship between the real
and imaginary parts of the Faraday angle as well as the Kerr
angle. Therefore, the imaginary part may be calibrated against the
real part. In this way, calibration of the system is performed as
a function of magnetic field in the same manner as samples of
interest without the addition of other optical components.
Utilizing the formalism developed in Sec.\,\ref{sec;AppGaAs}, the
real and imaginary part of the Faraday and Kerr angle data shown
in Fig.\,\ref{Cal2deg}(a-d) are simultaneously fit where the
mobility $\mu$, cyclotron mass m, 2D number density n, and sheet
resistance of the AR coating $R_\square$ are free parameters. The
B-field value of the resonance determines the mass m, the width of
the resonance determines the scattering rate characterized by the
mobility $\mu$, the overall strength of the response determines
the number density n, and deviations of the AR coating sheet
resistance R$_\square$ away from the ideal value tends to mix the
real and imaginary parts of the angles giving rise to anisotropies
in the resonant peaks.

The fit parameters are found to be the following: n=$2.0\times
10^{11}\,\text{cm}^{-2}$, $\mu=-1.19 \times 10^5 \text{cm}^2 /
\text{V\,sec}$ where the negative sign denotes the sign of the
carriers, m=0.0695\,m$_e$, and R=$1.01 \times \, \text{R}_{AR}$
where m$_e$ is the mass of an electron and $\text{R}_{AR}$ is the
sheet resistance for the perfect AR coating condition of GaAs
given by $Z_0/(n_{GaAs}-1)=\text{145}\,\Omega/\square$ where
$Z_0=\text{377}\,\Omega/\square$ is the impedance of free space
and $n_{GaAs}=3.6$ is the index of refraction for GaAs. The mass
is consistent with the known mass of
GaAs,\cite{GaAs-MassVsDoping1979} and the antireflection coating
is very close to ideal. The fits are overlayed on top of the
reflection and transmission data shown in
Fig.\,\ref{Cal2deg}(e,f). Essentially, all 6 curves are very well
reproduced with only two parameters, n and $\mu$. Together with
the polarizer calibration, the 2-DEG GaAs calibration indicates
that both the Kerr and Faraday experiments perform exceptionally
well.

This particular data set also illustrates the robustness of the
Kerr system. Instability associated with the two-laser system
occurred between +8 and $\sim$+5.5\,T appearing as a loss of power
and commensurate increase of relative detector noise in the Kerr
and reflection data sets. Although the reflection data was visibly
affected, the Kerr signal was essentially unaffected. The
reflection measurement is derived by the ratio of the signals from
two detectors (`Kerr detector' and `reference detector' in
Fig.\,\ref{figOptics}) and therefore susceptible to large spacial
mode fluctuations, but the Kerr signal is derived from the
modulated signal measured by a single detector.

\begin{figure*}[!t]
\includegraphics[scale=.7,clip=true, trim = 50 0 50 0]{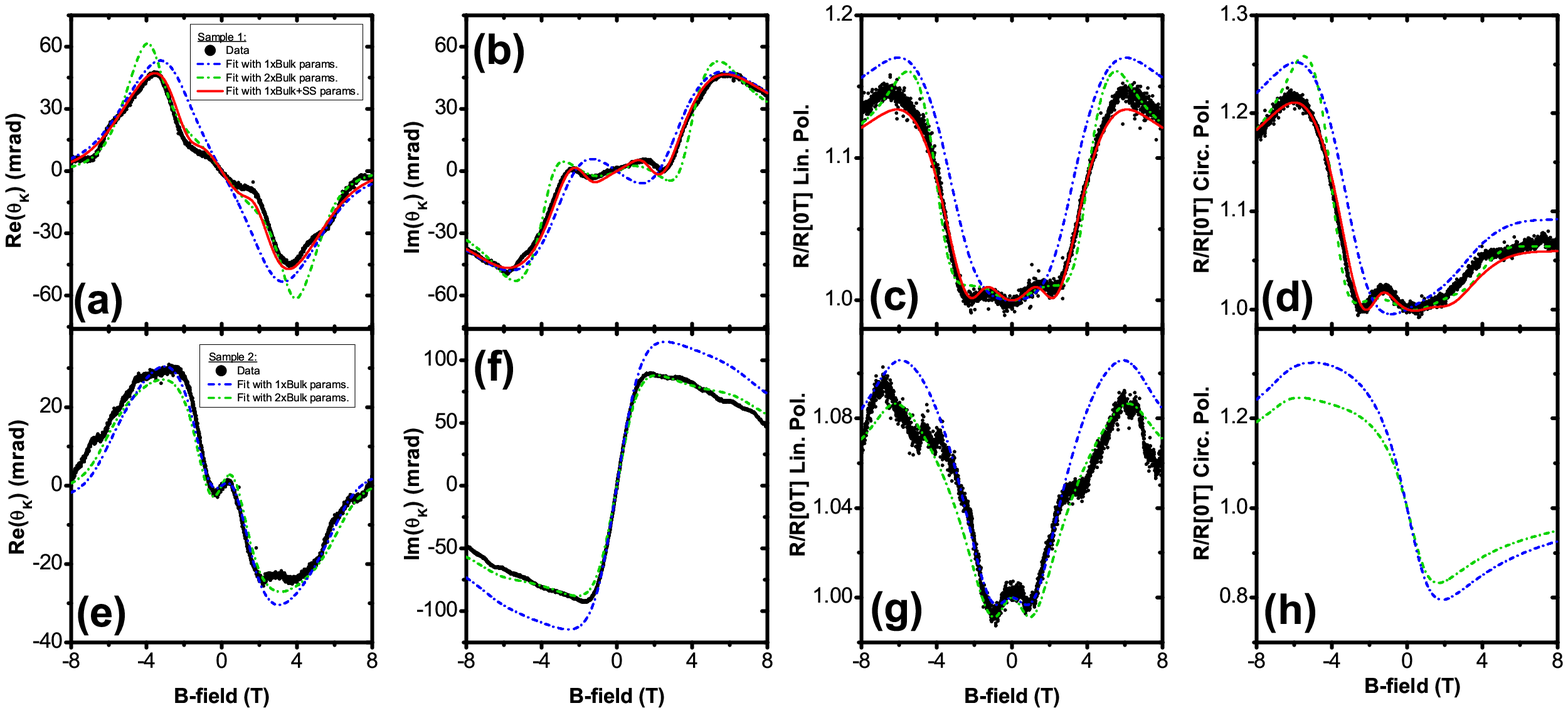}
\caption{\label{fig;alldata+fits}(color online) Kerr angle and
reflection of linearly polarized light normalized to the B=0\,T
value were measured at 10\,K and an optical frequency of
5.24\,meV. Reflection of nearly circularly polarized light
normalized to the B=0\,T value was a separate measurement
performed only on Sample 1. (a-d) Sample 1 data (black dots) is
depicted with several different fits superimposed. A homogeneous
bulk response can not describe the data (blue dash-dot). A
homogeneous bulk response with surface states (red solid) fit the
data. The weighted optical response of two crystallites with
different carrier density (green dash-dot) can reproduce the
salient features of the data. Slightly elliptically polarized
light, characterized by parameter $\beta$=0.13, is used in
calculations for Fig.\,\ref{fig;alldata+fits}(d). (e-h) Sample 2
data (black dots) show a very different response but was only
2\,mm away from Sample 1 in the as-grown crystal.  A homogeneous
bulk response (blue dash-dot) does not quite describe the data.
Addition of a surface state response does not improve the fits.  A
weighted response of two crystallites with different carrier
density (green dash-dot) more closely resembles the data. No
reflectivity data using circular polarized light was taken for
Sample 2. $\beta$ is assumed zero in calculations in
Fig.\,\ref{fig;alldata+fits}(h). }
\end{figure*}


\subsection{Bi$_2$Se$_3$ sample description}

Bulk crystals of Bi$_2$Se$_3$ were grown by varying the heating
conditions of the melt and stoichiometric ratio of bismuth and
selenium.\cite{butch_PRB2010} Surfaces were cleaved perpendicular
to he c-axis, and the backsides were cut and ground yielding a
total bulk thickness of about 0.5\,cm. Samples were mounted such
that the applied magnetic field was parallel to the c-axis and the
radiation propagation direction. The effective area measured was
defined by the optical spot size $\sim\diameter $1.0\,mm. During a
time span of about 10 seconds, the surfaces of the crystals were
cleaved using scotch tape in air and immediately inserted into a
vacuum of $\sim 10^{-7}$ Torr.

\section{Kerr and Reflection measurements on $\text{Bi}_2 \text{Se}_3$ and model description}
The Kerr angle and concurrent reflection measurements of linearly
polarized light are shown in Fig.\,\ref{fig;alldata+fits}(a-c) and
(e-g) for two samples both measured at 10\,K in fields up to 8\,T.
A separate measurement was performed on Sample 1 where the
waveplate in the rotator depicted in Fig.\,\ref{figOptics} was
statically set producing elliptically (nearly circularly)
polarized light. The reflection was then measured and normalized
to the B=0\,T value. The data is reported in
Fig.\,\ref{fig;alldata+fits}(d). $\beta$ characterizes the degree
of mixing between left and right circular polarization states via
$(1-\beta )\text{R}_+ + \beta \text{R}_-$ where R$_\pm$ is the
reflection produced by purely circularly polarized light. In this
measurement, $\beta$ is found to be 0.13.

To analyze the results, a model is constructed where the bulk is
considered to be a thick slab whose optical response is
characterized by a dielectric function consisting of a Drude
carrier response with an associated unscreened bulk plasma
frequency $\omega_{pb}$, bulk cyclotron mass $m_{cb}$, and bulk
scattering rate $\gamma_b$ along with the addition of two TO
phonon modes whose parameters have been previously
determined.\cite{sushkov_BiSe_2010} The surface state resides at
the interface whose admittance is modelled within a Drude model
with an associated 2-D surface plasma frequency $\omega_{ps}$,
surface cyclotron mass $m_{cs}$, and surface scattering rate
$\gamma_s$. Details of the model are discussed in
Sec.\,\ref{sec;AppBiSe}.

\section{Discussion}
\subsection{Analysis of the optical response}
Sample 1 data shown in Fig.\,\ref{fig;alldata+fits}(a-d) can not
be quantitatively nor qualitatively reproduced by a homogenous
bulk response with any arbitrary carrier concentration, cyclotron
mass, or scattering rate. The blue dash-dot curves in
Fig.\,\ref{fig;alldata+fits}(a-d) show a representative example of
line-shapes expected from a simple bulk response, calculated with
the specific parameters $\omega_{pb} = 420\, \text{cm}^{-1}$
(corresponding to a number density of $n_b \sim 3.2 \times 10^{17}
cm^{-3}$), $\gamma_b = 12\,\text{cm}^{-1}$, and $m_{cb}= .135\,
m_e$. Although the high field  Kerr data in
Fig.\,\ref{fig;alldata+fits}(a,b) is well described by this
particular set of parameters, strong qualitative deviations exist
at fields below $|B|\lesssim 4$\,T suggesting that another carrier
response with a much smaller cyclotron mass may be superimposed.
The reflection data in Fig.\,\ref{fig;alldata+fits}(c,d) also show
very different behavior at low fields $|B|\lesssim 4$\,T compared
to a bulk response. The small humps which appear at fields around
$\sim\pm1$\,T in Fig.\,\ref{fig;alldata+fits}(c) and $\sim -1$\,T
in Fig.\,\ref{fig;alldata+fits}(d) can not be qualitatively
reproduced with any value of bulk carrier concentration, cyclotron
mass, or scattering rate.

The addition of a surface state with parameters $\omega_{ps}=
41\,\text{cm}^{-1}$, $\gamma_s= 28\,\text{cm}^{-1}$, and $m_{cs}=
0.033\, m_e$ results in the red solid curves in
Fig.\,\ref{fig;alldata+fits}(a-d) which fit all the data
remarkably well. The surface state cyclotron mass corresponds to a
resonance at B=1.5\,T (for a frequency of 5.24\,meV). The
scattering rate associated with the surface is substantially
larger than that of the bulk, a result which might be expected
based upon Shubnikov de Haas measurements on similarly grown
samples\cite{butch_PRB2010} and linewidths estimated from ARPES
data.\cite{Xia-Hassan_ARPESBiSe_2009} Reasonable linefits may be
achieved with a surface state plasma frequency as low as
$\omega_{ps}= 20\,\text{cm}^{-1}$ although deviations from the
data begin to become more pronounced.

From the ARPES measured dispersion of the surface states (see
Eq.\,\ref{eq;ARPESns-ms}),\cite{Xia-Hassan_ARPESBiSe_2009} the
cyclotorn mass given by $\hbar k_f / v_f$ and number density given
by $\pi k_f^2 / (2 \pi)^2$ may be parameterized in terms of the
Fermi energy. A mass of 0.033\,m$_e$ corresponds to a Fermi energy
of $\sim$20\,meV above the Dirac point, a number density of
$10^{11}\,\text{cm}^{-2}$, and corresponding surface plasma
frequency of $1.3\,\text{cm}^{-1}$. The bulk plasma frequency
corresponds to a Fermi energy of approximately 20\,meV above the
bulk conduction band edge which is itself located $\sim 200$\,meV
above the Dirac point.

ARPES observations correspond to a large occupation number of the
surface states, $n_s\sim10^{12}\,\text{cm}^{-2}$.
\cite{Xia-Hassan_ARPESBiSe_2009} This amount of charge acting on
the bulk carriers would deplete the free carriers to a depth
$d=n_s/n_b \sim 100$\,nm (where $n_b$ is the bulk number density).
Since ARPES concurrently observed the bulk bands but the
measurement technique is sensitive to depths of only $\sim$1\,nm,
it must be assumed that there is a fixed neutralizing positive
surface charge coming from excess Se$^+$ or adsorbed ions, or
possibly an intrinsic effect. Therefore, band bending effects are
apparently small in vacuum cleaved crystals. However, for samples
exposed to ambient conditions, band bending may be more
severe\cite{analytis_SdH_PRB2010} possibly due to oxidation of the
surface.  This chemical doping moves the Dirac point toward the
chemical potential and leads to a depletion region in the bulk,
presumably with $d<n_s/n_b$.  Since the optical penetration depth
is much larger than 100\,nm at terahertz
frequencies,\cite{sushkov_BiSe_2010} an insulating depletion layer
has no significant optical response and the experiment sees the
surface carriers only as a modification to the bulk.

More significantly, the surface plasma frequency obtained from the
fits to the optical response are about a factor of 15 to 30 larger
than predicted from conventional transport theory based on the
ARPES measured surface state dispersion relation.  A possible
explanation is that the crystal has multiple surface layers within
the optical penetration depth, perhaps separated by interstitial
selenium, dividing the crystal into $\sim$10 thin flakes which are
optically very close but electronically decoupled. The cleave
process may favor the region with layers of flakes where these
interstitial selenium may be more prevalent offering a weaker link
between planes.

Given the unconventional spin-chiral nature of the eigenstates,
which imply spin currents perpendicular to the charge currents, an
unconventional matrix element for Landau level transitions may
also lead to enhancement of the spectral weight, a possibility
that needs further investigation.

As a possible alternative explanation, the structure observed at
low field in the optical response of Sample 1 may arise from
spatial variation of the bulk charge carrier concentration.
Consider the possibility that the optical spot covers a region
where two crystallites of different carrier concentrations exist.
Modelling this scenario, a simple weighted average of two
individual bulk responses without surface states is performed with
the following parameters: one bulk response weighted by 0.84 with
parameters $\omega_{pb1}= 334 \,\text{cm}^{-1}$,
$\gamma_{pb1}=6.5\,\text{cm}^{-1}$, and $m_{cb1}= 0.126$ with a
second bulk response weighted by (1-0.84) with parameters
$\omega_{pb2}=548\,\text{cm}^{-1}$, $\gamma_{b2}=
8.2\,\text{cm}^{-1}$, $m_{pb2}=0.148\, m_e$. The green dash-dot
curves in Fig.\,\ref{fig;alldata+fits}(a-d) show all of the
salient features of the data which was attributed to the surface
states. The difference between the two bulk masses is unexpected
for a simple parabolic band near a gapped band edge. The coarse
qualitative features of the data may be reproduced by choosing the
bulk masses to be the same with the other parameters chosen
similarly although the fits are substantially worse. Therefore,
even though the fit is not very good quantitatively, it is
possible that a distribution of carrier concentrations may be the
culprit giving rise to the observed line-shapes.

Perhaps lending more credence to the spatial variation of carrier
concentration argument is that a completely different response was
measured by a separate sample which was only 2\,mm away from
Sample 1 in the $\sim\diameter 1.5\,\text{cm}$ as-grown crystal.
The measurement results of Sample 2 are shown in
Fig.\,\ref{fig;alldata+fits}(e-g). A single bulk response
qualitatively describes all the features of the data, as depicted
by the blue dash-dot curves, whose parameters are $\omega_{pb} =
583\,\text{cm}^{-1}$, $\gamma_b = 7\,\text{cm}^{-1}$, and $m_{cb}=
0.145\, m_e$. $\beta$ is assumed zero in calculations in
Fig.\,\ref{fig;alldata+fits}(h). This is a much higher bulk plasma
frequency than found for Sample 1. Addition of a surface state
response does not improve the fits. However, modelling the bulk
carrier concentration variations as was done for Sample 1 where
one bulk response is weighted by 0.70 with parameters
$\omega_{pb1}= 583 \,\text{cm}^{-1}$, $\gamma_{b1}=
5.3\,\text{cm}^{-1}$, and $m_{cb1}= 0.145 \,m_e$ and a second bulk
response weighted by (1-0.70) with parameters
$\omega_{pb2}=250\,\text{cm}^{-1}$, $\gamma_{b2}=
15\,\text{cm}^{-1}$, $m_{cb2}=0.145\, m_e$ yields a good
quantitative fit to all three data sets as shown by the green
dash-dot curves in Fig.\,\ref{fig;alldata+fits}(e-g). A third
sample measured 2\,mm away from Sample 2 (and 4\,mm away from
Sample 1) yielded the same response as sample 2.

Controlling carrier concentrations to well within a factor of
$\sim 10$ over cm-sized crystals is very challenging. Using the
Kerr and reflectivity measurements, it is not possible to
determine whether such a response originates from a distribution
of carrier concentration in a single crystallite, or the presence
of two (or more) crystallites with different concentrations.

\subsection{Other methods to characterize surface states}

Using the Kerr measurement technique as a routine diagnostic of
sample carrier concentration variation is not feasible since it is
resource intensive. Since inhomogeneity is an important general
issue, a routine diagnostic tool is required. One method presently
being explored utilizes FTIR spectroscopic reflection
measurements. As representatively shown in the reflectivity
spectral measurements in figure 4 of
Ref.\,\onlinecite{sushkov_BiSe_2010}, a large dip occurs at the
plasma edge. The frequency location of the dip gives information
regarding the carrier concentration. Measurements on different
parts of a sample utilizing a small spot size
$\sim\diameter$0.5\,mm in the infrared spectral region is
currently being investigated as a diagnostic tool to quantify the
degree of carrier concentration variation.

Another method that may be favorable for observing surface states
presently being developed utilizes a gate. If the surface states
have a cyclotron mass and optical spectral weight commensurate
with the ARPES measured dispersion, the line-fitting procedure as
used for the above Kerr and reflection data for discerning the
surface state will most likely be extremely difficult with the
present signal-to-noise of the system, even for fairly small
surface scattering rates (see Sec.\,\ref{ExpectedKerrRfromARPES}
for details).

By gating, the surface state optical response can be modulated
with negligible effects produced by the bulk.  Any modulated
signal coming from the bulk would necessarily be due to the
screening layer after the depletion layer which is of thickness
$\sim 1/k_s$, the Thomas-Fermi screening length.  $k_s \sim
15\,\text{nm}$ at $n_b=10^{17}\,\text{cm}^{-3}$, a typical carrier
concentration.  The charge in this layer $n_d \sim n_b/k_s \sim
10^{11}\,\text{cm}^{-2}$  which is much less than the surface
state carrier density.  More important, however, is that $k_s$
does not depend on gate voltage and therefore is not modulated.
Secondly, by modulating a gate, signal-to-noise is expected to
increase by over an order of magnitude. Averaging between gate-on
and gate-off configurations in fixed field on relatively short
time scales eliminates magnetic field drift issues and minimizes
time drift issues. Thirdly, due to this large gain in sensitivity,
even very high surface scattering rates should be detectable.
Expected signals produced by gating, modelled in
Sec.\,\ref{sec;Kerr+gate}, are expected to be well within
experimental capability. Fourthly, the expected Kerr response from
the surface states for bulk plasma frequencies ranging from 200 to
600\,cm$^{-1}$ have qualitatively the same line-shapes. Even with
possible carrier inhomogeneities, the detection of surface states
is possible by modulating a gate with modest shifts of the surface
state Fermi energy. However, characterization of the surface
states will require reasonably homogeneous carrier concentrations.
Lastly, developing the technique is a necessary stepping stone for
future experiments: measurement of the half-quantized Hall
conductivity in topological insulators inside the gap created by a
magnetic layer.\cite{QiZhang_PRB2008}

\section{Conclusion}
Kerr measurements using heterodyne detection in the terahertz
regime are reported. Measurements of a 2-DEG GaAs heterostructure
prove the system works exceptionally well. Results on Bi$_2$Se$_3$
crystals show that the technique is sensitive to bulk carrier
concentration. For one particular sample, the Kerr and reflectance
data are very well described by including surface states. The
measured surface spectral weight $\omega_{ps}$ (with an associated
surface cyclotron mass $m_{cs}=0.033\,\text{m}_e$) is $\sim$20
times larger than expected for a single surface state based upon
ARPES measurements. Spatial variation of carrier concentration can
also qualitatively account for the data. Separation of bulk and
surface states may be accomplished in the Kerr geometry with
greater sensitivity by incorporating a gate, a proposed next step
to clearly separate the bulk and surface state effects.

\section{Acknowledgement}
This work was supported by the CNAM, NSF (DMR-0030112), and MRSEC
(DE-AC02-98CH10886). N. P. B. was supported by CNAM. We thank
Michael Fuhrer and Tudor Stanescu for insightful discussions.

\appendix
\section{\label{secAppendixMultilayer}Multilayer response analysis}
\subsection{General Notation}
The formalism for deriving the Fresnel reflection and transmission
amplitudes for a system consisting of multiple layers of
`dielectric' slabs with thin films at the interfaces are derived
for a general number of layers. The model will be applied to two
specific cases: a GaAs 2-DEG heterostructure with a metallic
antireflection coating and a bulk Bi$_2$Se$_3$ crystal with a
surface state.

Using Maxwell's equations for radiation at normal incidence,
boundary conditions on the electric and magnetic field may be
applied across any boundary between two dielectric slabs $i$ and
$i+1$ of thicknesses $db_i$ and $db_{i+1}$ with bulk complex
indices n$_i$ and n$_{i+1}$ and a metallic thin film at the
interface with a sheet conductivity of $\sigma_{i} \, df_i$ where
$df_i$ is the thickness of the film. The vector notation
(E$^r$,E$^l$) represents right and left travelling waves. The
electric field may be expressed in either a linear polarization
basis notated as (E$^r_x$,E$^l_x$) and (E$^r_y$,E$^l_y$) or in a
circular polarization basis as (E$^r_+$,E$^l_+$) and
(E$^r_-$,E$^l_-$). To derive the equations generally, we use the
notation (E$^r_{\alpha}$,E$^l_{\alpha}$) to represent one of the
polarization basis states y, x, +, or -.

The matrix $M_{i,\alpha}$ which transfers an arbitrary E-field
(E$^r_{\alpha}$,E$^l_{\alpha}$) across an interface consistent
with Maxwell's equations is given by the following:
\begin{equation*} \label{eqnMi}
\begin{tabular}{ll}
$M_{i,\alpha}$ &=  $ \frac{1}{2\,n_{i+1,\alpha}} \times$ \\
&$\begin{pmatrix}
    n_{i,\alpha}+n_{i+1,\alpha} - y_{i,\alpha}  &   -n_{i,\alpha}+n_{i+1,\alpha} - y_{i,\alpha}   \\
   -n_{i,\alpha}+n_{i+1,\alpha} + y_{i,\alpha}    &   n_{i,\alpha}+n_{i+1,\alpha} +
   y_{i,\alpha}
\end{pmatrix}$\\[1.0ex]
\end{tabular}
\end{equation*}
where the admittance of the thin film is $y_{i,\alpha}=Z_0
\sigma_{i,\alpha}\, df_i$ and $Z_0$ is the impedance of free
space.

The matrix $K_{i,\alpha}$ transfers an arbitrary oscillating
E-field (E$^r_{\alpha}$,E$^l_{\alpha}$) at frequency $\omega$
across a dielectric of thickness $db_i$ and complex index
$n_{i,\alpha}$ is given by the following:
\begin{equation*} \label{eqnKi}
\begin{tabular}{lcr}
$K_{i,\alpha}$ &= &$
\begin{pmatrix}
    e^{\imath \, n_{i,\alpha} \, db_i \, \omega/c}  &   0   \\
   0    &   e^{-\imath \, n_{i,\alpha} \, db_i \, \omega/c}
\end{pmatrix}$\\[1.0ex]
\end{tabular}
\end{equation*}

In general, an incident right travelling wave will be assumed
incident on the first interface with a resulting net reflected
travelling wave to the left from the first interface and a net
right travelling transmitted wave from the last interface. The
Fresnel transmission and reflection amplitudes for a stack of
dielectrics of m interfaces with thin metallic films are then
found by the relation:
\begin{equation*} \label{eqnr-t-arb}
\begin{tabular}{lcr}
$\begin{pmatrix} t_\alpha \\ 0 \end{pmatrix}$ &=&  $M_{m,\alpha}
\displaystyle\prod_{i=1}^{m-1}{K_{i+1,\alpha} M_{i,\alpha}}$ $\begin{pmatrix} 1 \\
r_\alpha
\end{pmatrix}$\\[1.0ex]
\end{tabular}
\end{equation*}

\begin{figure*}[!t]
\includegraphics[scale=.7,clip=true, trim = 50 0 50 0]{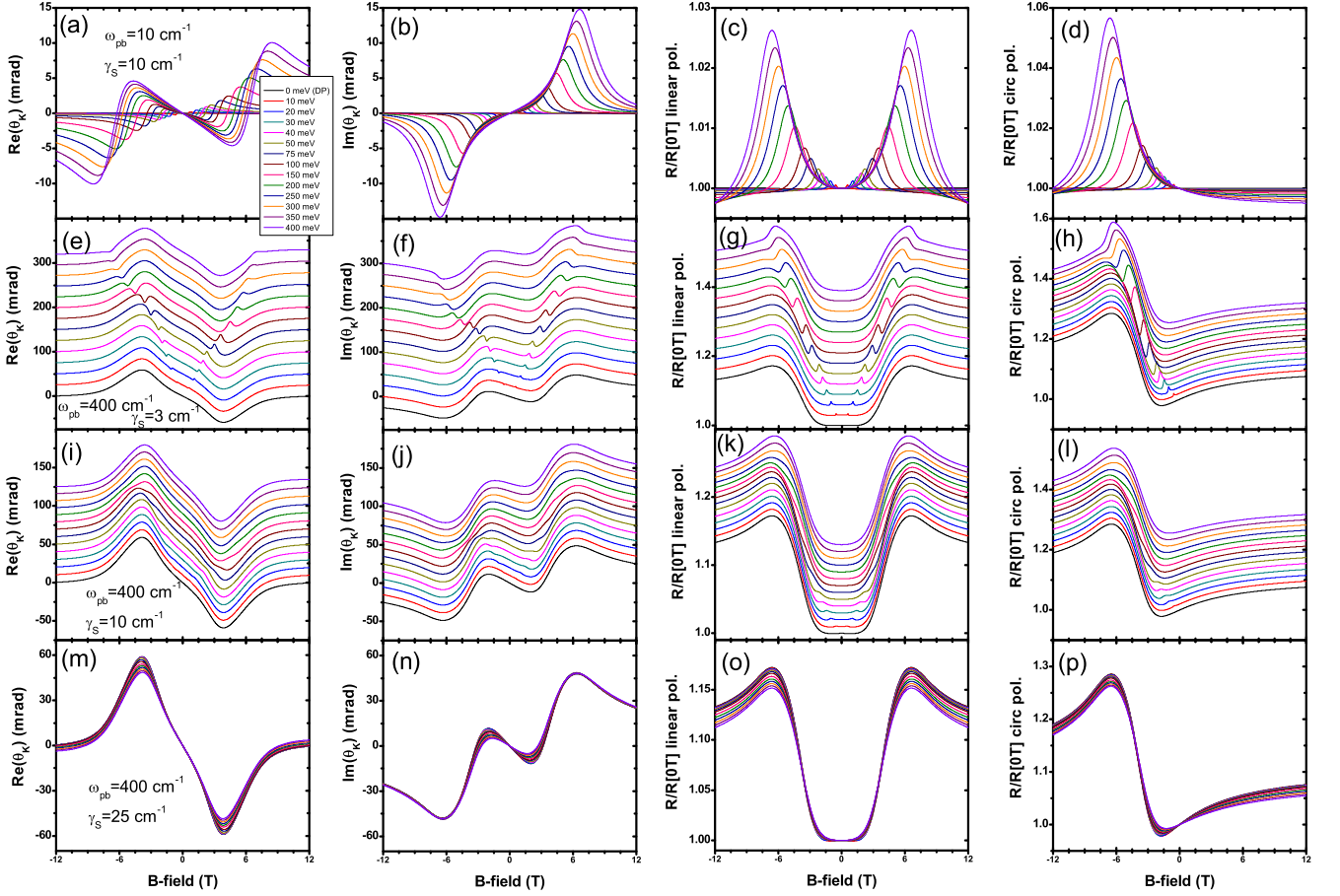}
\caption{\label{fig;SS-Bulk+SS}(color online) Calculated optical
response produced by the bulk properties and the expected surface
state response as measured by ARPES of Bi$_2$Se$_3$ as a function
of magnetic field for series of surface state Fermi energies
measured above the Dirac point as given in the legend: $E_F$=0,
10, 20, 30, 40, 50, 75, 100, 150, 200, 250, 300, 350, and 400 meV.
The Kerr angle as well as the reflectance (normalized to zero
field) in a pure circular or linear state are calculated. Each row
of graphs represents a different set of Drude parameters
describing the bulk and surface state response. The second and
third row graphs, (e-h) and (i-l), are offset for clarity such
that the Kerr angle is zero and R/R[0T] is unity in zero magnetic
field. The graphs with the Fermi energy $E_F=0$ correspond to the
Dirac point where there is no contribution from the surface state,
only bulk response. All graphs are calculated with a bulk mass
$m_{cb}=0.15\,m_e$ and a bulk scattering rate
$\gamma_b=10\,\text{cm}^{-1}$ and using the bulk phonon parameters
as measured in Ref.\,\onlinecite{sushkov_BiSe_2010}. (a-d) Very
low bulk density to demonstrate the response expected from the
surface states: $\omega_{pb}=10 \,\text{cm}^{-1}$, $\gamma_s=10\,
\text{cm}^{-1}$. (e-h) $\omega_{pb}=400 \,\text{cm}^{-1}$,
$\gamma_s=3 \, \text{cm}^{-1}$ (i-l) $\omega_{pb}=400
\,\text{cm}^{-1}$, $\gamma_s=10 \,\text{cm}^{-1}$. (m-p)
$\omega_{pb}=400 \,\text{cm}^{-1}$, $\gamma_s=25\,
\text{cm}^{-1}$.}
\end{figure*}

\subsection{GaAs 2-DEG}\label{sec;AppGaAs}

The above formalism can be applied to calculate the Kerr and
Faraday angle as well as the reflection and transmission for the
calibration sample. A GaAs 2-DEG is modelled as a thin metallic
sheet whose sheet conductance is described within a simple Drude
model located at the interface between vacuum and a GaAs
dielectric slab. The backside of the sample consists of an
interface between a GaAs dielectric slab and vacuum with an
antireflection coating, a thin metallic NiCr film, modelled within
a Drude model.  The antireflection coating is fully characterized
by the dc sheet resistance, $R_\square = 1/(\sigma \, df)$, since
the scattering rate of  NiCr is very large compared to the optical
frequency. The circular polarization basis is easiest to work with
in applied magnetic field.

The parameters are as follows: $n_{1,\pm}=1$, $n_{2,\pm}=3.6$,
$n_{3,\pm}=1$, $\omega = 5.24 \,\text{meV}$, $(Z_0 \sigma\,
df)_{1,\pm} = \omega_{ps1}^2 / (\gamma_1 - \imath (\omega \pm
\omega_{cs1}))$, $db_2=0.5\,\text{mm}$, and $(Z_0 \sigma\,
df)_{2,\pm} = Z_0/R_\square$ where $\omega_{cs1} = e B / (m c)$ is
the cyclotron frequency of the 2-DEG, $\gamma_1= e/(m \, \mu)$ is
the scattering rate of the 2-DEG, $\omega_{ps1} = 4 \pi n e^2/ (m
c)$ is the plasma frequency of the 2-DEG, m is the quasiparticle
mass, B is the applied magnetic field, e is the electronic charge,
c is the speed of light, $\mu$ is the mobility, and n is the 2-D
number density.

The Fresnel amplitudes $r_\pm$ and $t_\pm$ as a function of n,
$\mu$, R$_\square$, m, and B are then found by substituting the
above parameters into the following equations:
\begin{equation} \label{eqnr-t-circ}
\begin{tabular}{lcr}
$\begin{pmatrix} t_+ \\ 0 \end{pmatrix}$ &=&  $M_{2,+}\,
K_{2,+}\, M_{1,+}$ $\begin{pmatrix} 1 \\
r_+
\end{pmatrix}$\\
$\begin{pmatrix} t_- \\ 0 \end{pmatrix}$ &=&  $M_{2,-}\,
K_{2,-} \, M_{1,-}$ $\begin{pmatrix} 1 \\
r_-
\end{pmatrix}$\\[1.0ex]
\end{tabular}
\end{equation}

The reflection R and transmission T for unpolarized light,
reflection R$_\pm$ and transmission T$_\pm$ for circularly
polarized light, and the Faraday $\theta_F$ and Kerr $\theta_K$
angles are calculated from the Fresnel amplitudes as follows:
\begin{equation} \label{eqnRTFAKA}
\begin{tabular}{l}
R$_\pm$ = $r_\pm r^*_\pm$\\
T$_\pm$ = $t_\pm t^*_\pm$\\
R = $ (1/2) \, (R_+ + R_-)$  \\
T = $(1/2) \, (T_+ + T_-)$\\
$\tan(\theta_F)$  = $\imath (t_+ -
t_-)/(t_+ + t_-)$\\
$\tan(\theta_K)$ = $\imath (r_+ -
r_-)/(r_+ + r_-)$\\
\end{tabular}
\end{equation}
where `*' notates the complex conjugate.

\subsection{Bi$_2$Se$_3$}\label{sec;AppBiSe}
Bi$_2$Se$_3$ is modelled as a surface state Drude response on top
of a very thick bulk slab whose response consists of a free
carrier Drude term with two phonon resonances. The bulk phonon
parameters as well as the approximate cyclotron mass and
scattering rate have previously been
measured.\cite{sushkov_BiSe_2010}

The bulk dielectric function is given by the following:
\begin{equation}
\epsilon^{\pm} = - \frac{\omega_{pb}^2}{\omega (\omega \pm
\omega_{cb} -\imath \gamma_b)}
+\displaystyle\sum^2_{j=1}\frac{S_{j}}{\omega_{j}^2-\omega (\omega
- \imath \gamma_j)} + \epsilon_\infty \label{eps}
\end{equation}
where  $j$ enumerates the phonon oscillators, $\omega_{j}$ is the
TO phonon frequency, $S_j$ is the phonon spectral weight, a
subscript $b$ refers to bulk electrons, $\epsilon_\infty$ is the
dielectric constant at higher infrared frequencies, $\omega_{pb}$
is the unscreened plasma frequency for free electrons:
$\omega_{pb}^2=4\pi N_b e^2/m_b^*$, where $N_b$, $e$ and $m_b^*$
are concentration, charge, and an effective mass of electrons;
$\gamma_b$ and $\gamma_j$ are the damping rates; $\omega_{cb} = e
B / (m_{cb} c)$ is the bulk cyclotron frequency, where $B$ is
magnetic field, $m_{cb}$ is bulk cyclotron mass, and $c$ is the
speed of light.

The reported phonon parameters are $\omega_1=67$\,cm$^{-1}$,
$S_{1}=3.94\times 10^5$\,cm$^{-2}$, $\gamma_1= 5\,$cm$^{-1}$,
$\omega_2=134$\,cm$^{-1}$, $S_2=7056$\,cm$^{-2}$, $\gamma_2=
2$\,cm$^{-1}$, and $\epsilon_\infty =
25.6$.\cite{sushkov_BiSe_2010} The reported bulk cyclotron mass is
approximately $m_{cb} \approx 0.16 \, m_e$ where $m_e$ is the bare
electron mass and the approximate bulk scattering rate $\gamma_b
\approx 8 \,\text{cm}^{-1}$, although both are left as parameters
in the model. The relation $m_b^* = m_{cb}$ is assumed true, so
that $N_b$ (or equivalently the plasma frequency, $\omega_{pb}$)
is the third adjustable parameter associated with the bulk
response. With these values, $\epsilon_\pm$ is a function of
radiation frequency $\omega$ and applied magnetic field B, and the
bulk plasma frequency $\omega_{pb}$, mass $m_{cb}$, and scattering
rate $\gamma_b$.

Since the surface state penetration depth is $\sim$nm which is
much less than the optical wavelength for terahertz radiation, the
surface state may be modelled as a two dimensional sheet of
charge. The model of a 1\,cm thick Bi$_{2}$Se$_3$ slab in vacuum
with a Drude surface state response at the first interface are as
follows: $n_{1,\pm}=1$, $n_{2,\pm}=\sqrt{\epsilon_\pm}$,
$n_{3,\pm}=1$, $(Z_0 \sigma df)_{1,\pm} = \omega_{ps}^2 /
(\gamma_s - \imath (\omega \pm \omega_{cs}))$, $db_2=1
\,\text{cm}$, $(Z_0 \sigma \, df)_{2,\pm} = 0$ where $\omega_{cs}
= e B / (m_s c)$ and $\omega_{ps}= (4 \pi /c) n_{s} e^2 / m_s$.
Even for very low densities of bulk carriers, the properties of
the last interface do not effect the reflected optical response
due to the large attenuation within the thick bulk crystal.

Inserting the above values into Eqs.\,\ref{eqnr-t-circ} give the
reflection and transmission Fresnel amplitudes as a function of
$\omega$, B, $\omega_{pb}$, $\gamma_b$, $m_{cb}$, $\omega_{ps}$,
$\gamma_s$, and $m_{cs}$. The Fresnel amplitudes are also given in
Ref.\,\onlinecite{sushkov_BiSe_2010}, Eqs.\,4 to 6.

Digitizing and fitting the reported ARPES measured dispersion
relation for the surface state from
Ref.\,\onlinecite{Xia-Hassan_ARPESBiSe_2009} gives the following
functional form:
\begin{eqnarray}\label{eq;ARPESns-ms}
E_F = 2.02 k_f + 10.44 k_f^2
\end{eqnarray}
where the Fermi energy $E_F$ is measured in eV above the Dirac
point and $k_f$ is measured in $\AA^{-1}$. $n_{s}$ (or
$\omega_{ps}$), $m_{cs}$, $v_f$, and $k_f$ are fully determined by
the dispersion relation for any given Fermi energy.

Using the dispersion relation, R, T, R$_\pm$, T$_\pm$, $\theta_F$,
and $\theta_K$ are calculated from Eq.\,\ref{eqnRTFAKA} as a
function of $\omega$, B, $\omega_{pb}$, $\gamma_b$, $m_{cb}$,
$E_F$, and $\gamma_s$.

\section{Expected optical signatures of surface states based upon ARPES
measurements}
\subsection{Expected Kerr and Reflectance signals}\label{ExpectedKerrRfromARPES}

\begin{figure*}[!t]
\includegraphics[scale=.7,clip=true, trim = 50 0 50 0]{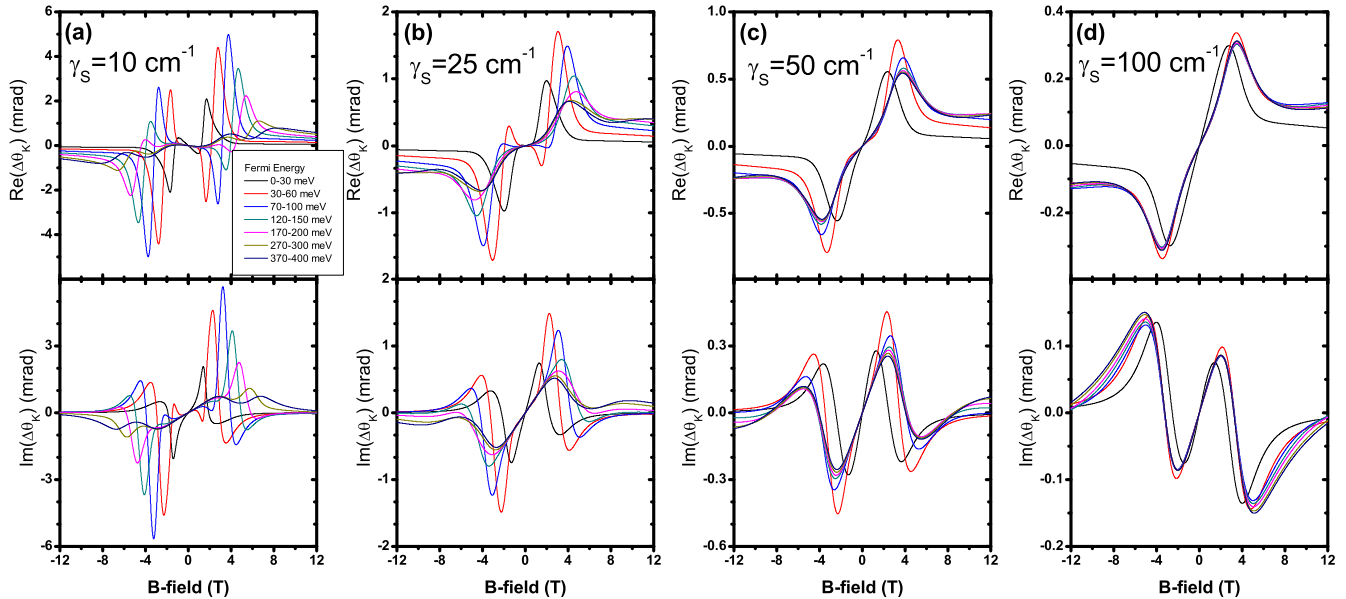}
\caption{\label{fig;gating}(color online) Calculated optical
response produced by a hypothetical gate voltage which shifts the
Fermi energy of the surface state by 30\,meV. The change in Kerr
angle produced by this shift at various Fermi energies for a
specific surface scattering rate is shown for each column. All
graphs are calculated using the bulk phonon parameters measured in
Ref.\,\onlinecite{sushkov_BiSe_2010} as well as the Drude bulk
parameters $\omega_{pb}=400 \,\text{cm}^{-1}$,
$\gamma_b=10\,\text{cm}^{-1}$, and $m_{cb}=0.15\,m_e$. The surface
state 2-D plasma frequency and mass are calculated based upon the
Fermi level and the measured ARPES dispersion
relation.\cite{Xia-Hassan_ARPESBiSe_2009} The surface scattering
rates are assumed to be (a)$\gamma_s=10 \text{cm}^{-1}$, (b)
$\gamma_s=25 \text{cm}^{-1}$, (c) $\gamma_s=50 \text{cm}^{-1}$,
and (d) $\gamma_s=100 \text{cm}^{-1}$. }
\end{figure*}

Figure \ref{fig;SS-Bulk+SS} shows calculated results for the
expected Kerr and reflection signals, both from linearly and
circularly polarized light, as a function of magnetic field for
various Fermi energies ranging from 0 to 400\,meV above the Dirac
point. All graphs in Fig.\,\ref{fig;SS-Bulk+SS} are calculated
assuming a bulk cyclotron mass value of $m_{cb}=0.15\,m_e$ and
bulk scattering rate $\gamma_b=10\,\text{cm}^{-1}$. The surface
state plasma frequency $\omega_{ps}$ and cyclotron mass $m_{cs}$
are determined by the ARPES measured dispersion
(Eq.\,\ref{eq;ARPESns-ms}) for a specific surface Fermi energy.
Each row of graphs are calculated with a different bulk plasma
frequency $\omega_{pb}$ and surface scattering rate $\gamma_s$.

The optical response for a pure surface state with
$\gamma_s=10\,\text{cm}^{-1}$ is shown in
Fig.\,\ref{fig;SS-Bulk+SS}(a-d) (the response of the bulk is
negligible for $\omega_{pb}= 10\,\text{cm}^{-1}$). All signals are
well within the capability of the Kerr measurements. Note that
Re($\theta_K$) and Im($\theta_K$) show antiresonant- and
resonant-like behavior. The cyclotron mass and strength of the
response increase with Fermi energy. The reflection in purely
circularly polarized light depicted in
Fig.\,\ref{fig;SS-Bulk+SS}(d) exhibits a hump in the cyclotron
resonant active mode. The reflection in linear polarization
depicted in Fig.\,\ref{fig;SS-Bulk+SS}(c) (composed of equal
amounts of left and right circularly polarized light) shows two
humps since there is an active cyclotron mode in positive and
negative fields. The qualitative structure of the Kerr and
reflection signals is instructive since they appear superimposed
on larger bulk responses at larger values of $\omega_{pb}$.

Row 2 of figures, namely Fig.\,\ref{fig;SS-Bulk+SS}(e-h), show the
optical response expected with a bulk plasma frequency of
$\omega_{pb}=400\,\text{cm}^{-1}$ and surface scattering rate
$\gamma_s=3\,\text{cm}^{-1}$. The bulk is depicted by the black
curves since no carriers are in the surface state at the Dirac
point, $E_f=0$. The curves are offset for clarity: in zero
magnetic field, the Kerr angle is always zero and the normalized
reflection is always equal to unity. With such a low surface
scattering rate, the sharpness of the surface state response is
seen clearly superimposed upon the bulk response.

The next row of figures, Fig.\,\ref{fig;SS-Bulk+SS}(i-l), show the
effect of increasing the surface scattering rate to
$\gamma_s=10\,\text{cm}^{-1}$. Again, the curves are offset for
clarity. Although the effect of surface states is discernable in
many of the curves, fitting the lineshapes with the current
signal-to-noise of the measurement system becomes problematic.

The last row of figures, Fig.\,\ref{fig;SS-Bulk+SS}(m-p), show the
effect of increasing $\gamma_s$ to $25\,\text{cm}^{-1}$. Fitting
lineshapes on this level to discern the effects produced by
surface states is not conducive to the current measurement scheme.

\subsection{Expected Kerr signals with application of a gate}\label{sec;Kerr+gate}
A gate which concurrently allows ample throughput optical power
while inducing negligible Kerr or Faraday rotation has been
accomplished.  High resistivity silicon substrates with a 300\,nm
film of SiO$_2$ were ion implanted with peak concentrations
$>10^{19}/\text{cm}^3$ resulting in a density profile thickness of
$\sim 0.3\,\mu\text{m}$. Characterization of these devices show
very small Faraday rotation up to 8\,T, excellent optical
throughput power, no carrier freeze out, and large break down
voltages. The devices are currently being used to gate large CVD
grown graphene in similar magneto-optical experiments.

Estimates of the shift in Fermi energy of the surface state
associated with the application of electric fields commensurate
with the typical breakdown voltages of SiO$_2$ depends upon the
Fermi energy. For currently available samples the surface state
charge density is sufficiently large that it dominates the charge
transfer to the sample due to the gate.  As a result the band
offset is reasonably given by the change in the Fermi energy due
to the charge transfer to the surface states.  The depletion depth
in the bulk is then given to a good approximation by $d^2=2 U_0
\epsilon_0/e^2 n_b$ (where $U_0$ is the energy above the Dirac
point and $\epsilon_0$ is the bulk static dielectric constant).
Therefore, the net charge which one can add or subtract from the
surface state without breakdown of the insulator is very nearly
equal to the maximum allowable charge transfer with the
application of a gate. Using a maximum applicable electric field
value of $3.3\times10^8$ V/m for SiO$_2$ is equivalent to a net
change in surface charge density of
$7\times10^{12}\,\text{cm}^{-2}$. Based upon the ARPES measured
dispersion, this gives a shift of Fermi energy  $\Delta E_f
\approx$90\,meV at $E_f$=0, $\Delta E_f \approx$30\,meV at
$E_f$=200\,meV, and $\Delta E_f \approx$15\,meV at $E_f$=350\,meV.
Since the gate voltage can be reverse biased as well, roughly
twice these changes can be realized. The resulting depletion depth
decouples the surface states, but is small enough that the
insulating depletion layer is optically invisible at THz
frequencies.

Based upon these estimates, the application of a gate is chosen to
produce a change in Fermi energy of $\Delta E_f=30$\,meV.
Fig.\,\ref{fig;gating} shows the expected Kerr signals predicted
by the ARPES measured surface state properties with bulk
parameters of $\omega_{pb}=400 \,\text{cm}^{-1}$,
$\gamma_b=10\,\text{cm}^{-1}$, and $m_{cb}=0.15\,m_e$. Each graph
shows Fermi level shifts of 0-30\,meV, 30-60\,meV, 70-100\,meV,
120-150\,meV, 170-200\,meV, 270-300\,meV, and 370-400\,meV. Graphs
(a-d) show the effects of increasing the surface scattering rate
from $10\,\text{cm}^{-1}$ to $100\,\text{cm}^{-1}$. Reflectance
measurements are not very sensitive, showing that R/R[0\,T]
$\lesssim$0.5\% for $\gamma_s=25\,\text{cm}^{-1}$ and
$\lesssim$0.1\% for $\gamma_s=50\,\text{cm}^{-1}$.

Changes in Kerr angle at the level of 0.1\,mrad is very reasonably
obtainable in this configuration. It should be noted that the
signals are maintained at these general levels even for higher
bulk plasma frequencies up to 600\,cm$^{-1}$.

\bibliography{KerrBiSev1}

%

\end{document}